%% file: paper_arxiv_2.tex
\documentclass{iopart}
\usepackage{iopams}
\usepackage{color}
\usepackage{graphicx}
\usepackage{braket}
\usepackage[bookmarks=false]{hyperref}

\definecolor{darkred}{rgb}{0.6, 0, 0}
\definecolor{darkgreen}{rgb}{0, 0.45, 0}
\definecolor{darkblue}{rgb}{0, 0, 0.6}
\definecolor{darkpurple}{rgb}{0.4, 0, 0.4}

\hypersetup{colorlinks=true, breaklinks=true, 
linkcolor=darkblue, citecolor=darkgreen}

\newcommand{\hint}{\ensuremath{\hat{H}_{\mathrm{int}}}}
\newcommand{\aop}{\ensuremath{\hat{a}}}
\newcommand{\adag}{\ensuremath{\hat{a}^{\dagger}}}
\newcommand{\bop}{\ensuremath{\hat{b}}}
\newcommand{\bdag}{\ensuremath{\hat{b}^{\dagger}}}
\newcommand{\cop}{\ensuremath{\hat{c}}}
\newcommand{\cdag}{\ensuremath{\hat{c}^{\dagger}}}

\newcommand{\myint}[1]{\ensuremath{\int\rmd #1}}
\newcommand{\lrsqrt}[1]{\ensuremath{\left(#1\right)^{1/2}}}
\newcommand{\ezero}{\ensuremath{\varepsilon_0}}
\newcommand{\chitwo}{\ensuremath{\chi^{(2)}}}
\newcommand{\hc}{\ensuremath{\mathrm{h.c.}}}

\begin{document}
%
%
\title[From quantum pulse gate to quantum pulse shaper]
{
	From quantum pulse gate to quantum pulse shaper -- engineered 
	frequency conversion in nonlinear optical waveguides
	}

\author{
	Benjamin Brecht$^{1,\dagger}$\footnotetext[1]{Author to whom any
	correspondence should be addressed}, Andreas Eckstein$^{2,1}$, 
	Andreas Christ$^{1,2}$, Hubertus Suche$^1$, and Christine 
	Silberhorn$^{1,2}$
	}

\address{
	$^1$ Applied Physics, University of Paderborn, 
	Warbuger Strasse 100, 33098 Paderborn, Germany
	}

\address{
	$^2$ Max Planck Institute for the Science of Light, 
	G\"unther-Scharowsky-Strasse 1 / Building 24, 91058 
	Erlangen, Germany
	}

\ead{benjamin.brecht@uni-paderborn.de}

\begin{abstract}
	Full control over the spatio-temporal structure of quantum
	states of light is an
	important goal in quantum optics, to generate for instance single-mode
	quantum pulses or to encode information on multiple modes, enhancing
	channel capacities. Quantum light pulses feature an inherent, 
	rich spectral broadband-mode structure. In recent years, exploring the 
	use of integrated optics as well as source-engineering has led to a deep 
	understanding of the pulse-mode structure of guided quantum states of 
	light. In addition, several groups have started to investigate the 
	manipulation of quantum states by means of single-photon frequency 
	conversion. In this paper we explore new routes towards complete control
	of the inherent pulse-modes	of ultrafast pulsed quantum states by employing 
	specifically designed nonlinear	waveguides with adapted dispersion 
	properties. Starting from our recently proposed quantum pulse gate (QPG)
	we further generalize the concept of spatio-spectral engineering
	for arbitrary $\chitwo$-based quantum processes.
	We analyse the sum-frequency generation based QPG
	and introduce the difference-frequency generation based
	quantum pulse shaper (QPS). Together, these versatile and robust integrated
	optics devices allow for arbitrary manipulations of the pulse-mode 
	structure of ultrafast pulsed quantum states. The QPG 
	can be  utilized to select an arbitrary pulse mode from a 
	multimode input state, whereas the QPS enables the 
	generation of specific pulse modes from an input wavepacket with 
	Gaussian-shaped spectrum.
\end{abstract}


\maketitle

%
%
\tableofcontents
\input{part_1_arxiv_2}
\input{part_2_arxiv_2}
\input{part_3_arxiv_2}
\input{part_4_arxiv_2}
\section*{References}
\bibliography{/Users/bbrecht/work/tex_stuff/bibliography}
\end{document}

%% file: part_1_arxiv_2.tex
\section{Motivation}
Ultrafast pulsed quantum states of light play an increasingly important role in 
quantum information and quantum communication as they allow for efficient 
network synchronization and high data transmission rates. In general they feature 
 a rich spectral mode structure, which is most naturally described
in a broadband pulse-mode basis. This is not a new result in either  
classical or quantum optics \cite{Titulaer1966}. For classical states all 
 basis sets are 
formally equivalent and no specific choice can be distinguished. In contrast, it has been shown that pulsed quantum
states of light exhibit an inherent pulse-mode structure, which is solely
determined by their generation process \cite{Martinelli2003}. Different kinds of
applications require specifically tailored pulsed quantum states, be it 
single-mode states for linear optical quantum computation \cite{Knill2001} or
multimode states for high-capacity quantum information encoding. Thus, a
thorough understanding of the spatio-spectral modal structure of ultrafast
quantum states as well as the ability to exercise full control over that 
structure is an important goal in today's quantum optical research.

 In this
paper we investigate the potential of engineered nonlinear waveguides for the
manipulation of pulsed quantum states which cannot be achieved within the
framework of linear optics. Special emphasis is put on an accurate 
description of the $\chitwo$-process inside the guide, which takes
into account rigorously the spatial and spectral degrees of freedom. Thus quantitative measures can be
derived  for the efficiency of practical 
quantum optical devices. 

The paper is organized as follows. In section \ref{sec:introduction}, we
review the state-of-the-art methods of generating ultrafast pulsed quantum states and  manipulating
 their inherent pulse-mode structure in bulk crystals and 
waveguides. We briefly discuss the latest developments and introduce new
ideas by combining dispersion engineering techniques, which have become  established by now for photon-pair preparation, 
with current methods of state 
manipulation utilizing $\chitwo$-nonlinearities. In this context we analyse
the experimental implementation of our recently proposed
quantum pulse gates (QPG) 
\cite{Eckstein2010} and extend the formalism  further by presenting 
 the concept of a quantum pulse
shaper (QPS). In sections \ref{sec:transformations} and \ref{sec:hamiltonian}
we develop a theoretical framework for our devices. We start with the linear
operator transformations for sum- and difference-frequency generation and
derive the interaction Hamiltonian of these processes considering spatial and
temporal degrees of freedom. Our analysis results in a completely
quantitative model. Section \ref{sec:application} is dedicated to merging
the derived theoretical framework with dispersion engineering methods known from
 state preparation, thus paving
the way to real-world applications, the performance of which is investigated
in section \ref{sec:performance}. Here we introduce realistic experimental
parameters for our waveguide devices and demonstrate the capability to fully
control the pulse-mode structure of ultrafast pulsed quantum states of light. 
Finally, in section \ref{sec:conclusion} we highlight the most important
results of this work, and end with an outlook on the use of QPG and QPS in 
continuous variable quantum information processing.
\section{Introduction}
\label{sec:introduction}
In recent years different approaches have been introduced to prepare and 
manipulate
ultrafast pulsed quantum states of light. 
One of the  most common sources for the generation of photonics quantum states is 
parametric downconversion (PDC) in nonlinear
crystals. This is mainly due to the rather simple experimental 
implementation
of PDC sources and their ability to achieve high photon-pair generation
rates. When pumped by ultrafast pulses, PDC processes generate
pulsed bi-photons with broad spectra. However, these states are
usually highly correlated due to the constraints imposed by energy and 
momentum conservation \cite{Keller1997, Grice2001}. 
Hence photon pairs are typically emitted
into many inter- and intra-correlated spatial-spectral 
modes,
the exact structure of which can be retrieved by applying a Schmidt 
decomposition to the biphoton amplitude distribution \cite{Law2000}. Upon
detection of one of the photons, the other one is projected 
onto a mixed state of all possible modes, rendering it ill-suited for 
linear optical quantum computation applications \cite{URen2005}. 
The common way for
overcoming this limitation has been narrowband spectral filtering to
force the photons into one optical mode \cite{Riedmatten2003, 
Kaltenbaek2006}. However, this approach prohibitively lowers the photon 
generation rate as most of the generated signal is lost in the
filtering process. It is thus
not feasible for large scale quantum information applications 
\cite{URen2005}. In addition,
only in the limit of infinitely narrow filtering one monochromatic, 
temporally de-localized
mode is selected, and the photon's pulse characteristic is lost. 
\subsection{Preparation of ultrafast pulsed quantum states with waveguides}
Only recently two new developments have made it possible to tackle
the aforementioned problems. The use of integrated waveguide sources 
has a major impact on the structure of PDC photon pair states.
In a bulk crystal the generated photons are emitted at the natural 
phasematching angles. This poses two problems: Firstly, the collection of 
the pair photons is
experimentally challenging and typically inefficient. 
Secondly, the pump field always 
couples to an infinite number of spectral-spatially correlated modes, and
thus the probability to create a photon pair in one distinct mode becomes
very low. In contrast, the emission in nonlinear waveguides is 
restricted to a well-defined set of discrete spatial modes defined by the
waveguide, ideally allowing only the propagation of one individual mode in a single-mode
wave-guide. \footnote{Here we do only consider guided modes neglecting any contributions,
which could be present due to phase-matched substrate modes. This simplification is justified,
because the continuous distribution of substrate modes can be easily filtered out by spectral or 
spatial filters.}
It turns out that the probability of
generating a photon pair in a distinct spatial mode is enhanced by several
orders of magnitude \cite{Tanzilli2001, Fiorentino2007}, since the total 
number of allowed modes is dramatically reduced inside the waveguide.
Moreover it also leads to an effective 
decoupling of the spatial from the spectral degree of freedom, since any 
spatial-spectral correlation necessitates more than one spatial mode. Even if 
other spatial modes apart from the ground mode are guided in the waveguide, 
modal waveguide dispersion usually ensures that both photons in those 
modes are created at different frequencies. Thus they can readily be removed 
by applying broadband spectral filters on the output state 
\cite{Mosley2009}.

The second step on the way to achieve complete control over the modal 
structure of
the generated quantum states of light is spectral source-engineering. 
It has been proposed that, by choosing adapted dispersion properties, 
photon pair generation can be tailored such that
signal and idler are emitted into one single spectral pulse mode each
\cite{URen2005, Garay-Palmett2009}. Later this has been experimentally 
demonstrated for bulk crystal sources \cite{Mosley2008, Wong2008, Shi2008} 
and photonic crystal fibre sources \cite{Cohen2009, Rarity2009, Soeller2010}. 
Only recently this has been realized in a waveguided parametric 
downconversion source in a KTiOPO$_4$ crystal \cite{Eckstein2010-2}. 
In this setup, the use of a waveguiding structure has led to an unprecedented
brightness for sources of separable photon-pair states. Note that narrowband
spectral filtering is not necessary with these sources as the generation 
process itself only allows one single spectral pulse mode.
Thus, the generated photon pairs are genuine quantum pulses, and are 
completely separable, spectrally as well as spatially.
\subsection{Manipulation of the pulse-mode structure of ultrafast 
quantum states}
Up to now, the research on manipulation of pulsed quantum states has mostly 
been focussed on shifting
their central frequency. It has been shown that sum-frequency generation 
(SFG) of single photons, in combination with subsequent photodetection, 
surpasses the efficiency of direct detection of near infrared single photons 
\cite{Roussev2004, VanDevender2004, Langrock2005, Albota2010}. Additionally 
SFG has been proven to conserve the quantum characteristics of the input
photon \cite{Huang1992,Tanzilli2005, VanDevender2007} and it has already
been utilized to implement measurement schemes with very high timing 
resolution, which overcomes long integration times of current 
single-photon detectors \cite{Wong2008}. Only last year, SFG has been 
demonstrated for single-photon Fock states \cite{Rakher2010}. Note that
recently also four-wave mixing in photonic crystal fibres has been
employed to demonstrate coherent frequency translation of single photons
\cite{McGuinness2010}. This highlights the broad interest and the numerous
application possibilities for these techniques.
With more and more single-photon sources available in the visible 
range, difference-frequency generation (DFG) has now also attracted 
increasing interest. Recent experiments employ DFG to 
implement wavelength interfaces for quantum networks \cite{Koshino2009, 
Ding2010, Takesue2010}, which equivalently to the SFG process preserve the quantum
characteristics of the input state. 
 
Despite this considerable progress, the generation of ultrafast pulsed quantum 
states with a specific pulse-mode structure, be it the number of excited 
modes or their shape, has not been explored, yet. Although the rich inherent
mode structure of ultrafast optical quantum states is well-known, before the
QPG \cite{Eckstein2010} there has been no feasible
way of controlling and manipulating the different modes separately.

\subsection{Quantum Pulse Gate and Quantum Pulse Shaper}

\begin{figure}
	\includegraphics[width=\linewidth]{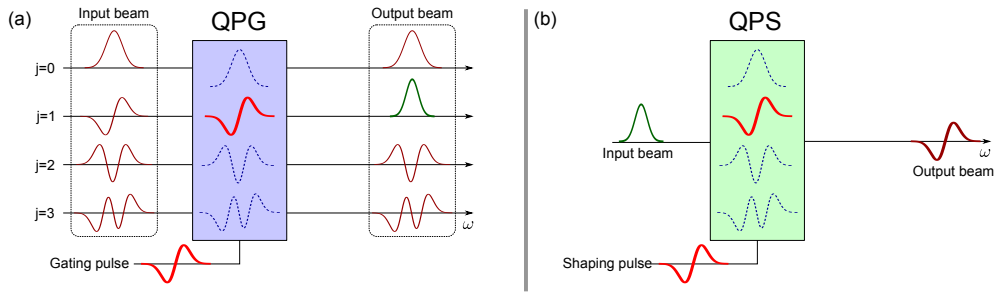}
	\caption{Schematics of (a) a Quantum Pulse Gate (QPG) and (b) a
	Quantum Pulse Shaper (QPS). 
	The QPG is based on engineered ultrafast sum-frequency generation in 
	nonlinear optical waveguides. By shaping an ultrafast gating pulse one
 	specific pulse mode from a pulsed multimode input state is selected and
	shifted to another frequency. Then it can easily be split off while
	leaving the rest of the state untouched.
	The QPS is based on engineered ultrafast difference-frequency generation
	in nonlinear optical waveguides. An input pulse mode can be converted 
	into an arbitrary output pulse mode by nonlinear interaction with an 
	ultrafast shaping pulse. The output pulse mode's shape is given by the
	mode of the shaping pulse.}
	\label{figure_1}
\end{figure}

We now combine the findings from the field of quantum-state generation with
the techniques from state manipulation. Applying source-engineering to
frequency conversion reveals fascinating possibilities to achieve the 
desired goal of complete control over the pulse-mode structure of ultrafast
quantum states.
In \cite{Eckstein2010} we have already proposed a quantum pulse gate (QPG),
a device based on engineered ultrafast SFG
in nonlinear waveguides. This device enables us to address different inherent 
pulse modes of an ultrafast pulsed quantum state of light individually
as illustrated in figure 
\ref{figure_1} (a). We would like to highlight that the QPG operation does 
not have any impact on 
the residual pulse-mode structure. This sets it apart from other experiments
which focus on a direct manipulation of the spectral broadband-mode structure 
of ultrafast pulsed quantum states and employ pulse shaping of photon-pair states
\cite{Dayan2004, Peer2005, Dayan2007}.
This alternative
approach also leads to highly interesting results for entanglement based 
applications. Still, the manipulation is not pulse-mode sensitive in the
sense of accessing and separating out a single-mode quantum state with 
specific temporal profile. In contrast, the QPG achieves mode selection by 
shaping an ultrafast, coherent gating pulse instead of the pulsed quantum 
state. The addressed mode is converted to the sum-frequency of input pulse
and gating pulse and is thus easily accessible. In addition, different 
orthogonal pulse modes can be interconverted into each other, rendering
interference between them possible. 
In this paper we elaborate on the QPG concept and come up with another fundamental device,
the quantum pulse shaper (QPS) based 
on engineered ultrafast DFG. While the QPG 
addresses single pulse modes, the QPS enables us to convert an 
input quantum state with Gaussian-shaped spectrum into a single-mode quantum state with arbitrary
shape (see figure \ref{figure_1} (b)). Here, an arbitrarily chosen pulse form
of the coherent shaping pulse defines the output pulse mode. 
We would like to mention that a similar idea
of shaping quantum pulses by means of frequency conversion with dispersion matching
has been proposed in
 \cite{Kielpinski2010}. In contrast to this earlier work we put special emphasis on 
the engineering of the dispersion characteristics of the used non-linear medium,
such that single-mode operation can be ensured avoiding the insertion of any unwanted
vacuum contributions.

Using QPG and QPS, pulsed quantum states can be generated and selected with unit efficiency in
arbitrary pulse forms and encoding of quantum information in broadband
mode basis and the successive read-out become possible. Therefore QPG and QPS 
will enable the
implementation of quantum communication protocols, which exploit the rich
pulse-mode structure of ultrafast states. 

%% file: part_2_arxiv_2.tex
\section{Linear transformations for SFG and DFG in comparison with PDC}
\label{sec:transformations}
In this section we qualitatively discuss the nonlinear three-wave mixing
processes SFG, DFG and PDC, highlighting their formal similarities as well
as examining their differences. In such a three-wave mixing process, three 
electrical fields interact inside a nonlinear medium, and the interaction 
Hamiltonian in the rotating-wave approximation is of the form
\begin{equation}
	\hint\propto\chitwo\myint{^3r}\hat{E}^{(+)}_\mathrm{a}(\vec{r}, t)
	\hat{E}^{(-)}_\mathrm{b}(\vec{r}, t)\hat{E}^{(-)}_\mathrm{c}
	(\vec{r}, t)+\hc
	\label{eq:basic_hamiltonian_efields}
\end{equation}
The $\hat{E}^{(+)}_i(\vec{r}, t)$ describe the positive frequency parts of the
interacting electric fields, $\chitwo$ is the second order nonlinearity
of the medium. In PDC
and single-photon SFG and DFG, two of the three fields are 
generally considered quantum mechanically. The remaining field is a bright, 
immutable pump field which can be treated classically. In this case the 
interaction Hamiltonian becomes bilinear and Heisenberg's equation of motion 
yields linear input-output transformations for the creation and annihilation
operators. Depending on which of the three fields is defined as pump, one can
distinguish two flavours of processes which are characterized by different
linear operator transformations. 
This can be derived when considering a single-mode approximation to 
equation \ref{eq:basic_hamiltonian_efields}:
\begin{equation}
	\hint\propto\aop\bdag\cdag + \adag\bop\cop. 
	\label{eq:basic_hamiltonian_operators}
\end{equation}
Firstly we assume that field $E_\mathrm{a}(\vec{r}, t)$ is the classical coherent 
pump field. We insert its classical amplitude $\alpha$ into the above equation
and find
\begin{equation}
	\hint\propto\alpha\:\bdag\cdag + \alpha^*\:\bop\cop.
\end{equation}
The resulting operator formally corresponds to a two-mode squeezing
operator (compare e.g. \cite{Barnett1997}), which means that this case describes 
PDC. Depending on the pump power of the bright field, either the photon 
pair characteristics (low power regime) or the squeezer characteristics 
(high power regime) dominate the PDC output state. 
The linear transformations between input and output operators evaluate to
\begin{eqnarray}
	\bop \rightarrow &\cosh(\zeta)\bop - \sinh(\zeta)\cdag,\\
	\cop \rightarrow -&\sinh(\zeta)\bdag + \cosh(\zeta)\cop,
\end{eqnarray}
where the parameter $\zeta$ depends on the pump power and is related to the
amount of squeezing in the generated pair state. This is discussed, for
instance, in \cite{Loudon2000}.

We find the other flavour of $\chitwo$ processes by assuming that field
$E_\mathrm{b}(\vec{r}, t)$ corresponds to the pump field. 
We substitute its 
classical amplitude $\beta$ in \eref{eq:basic_hamiltonian_operators} and obtain
\begin{equation}
	\hint\propto \beta\:\aop\cdag + \beta^*\:\adag\cop.
	\label{eq:intro_beamsplitter}
\end{equation}
This expression is formally equivalent to an optical beamsplitter 
Hamiltonian and we can use the well-known beamsplitter input/output
transformations for the operators $\aop$ and $\cop$.
\begin{eqnarray}
	\aop \rightarrow &\cos(\theta)\aop - \rmi\sin(\theta)\cop,\\
	\cop \rightarrow -&\rmi\sin(\theta)\aop + \cos(\theta)\cop.
\end{eqnarray}

\begin{figure}
	\includegraphics[width=\linewidth]{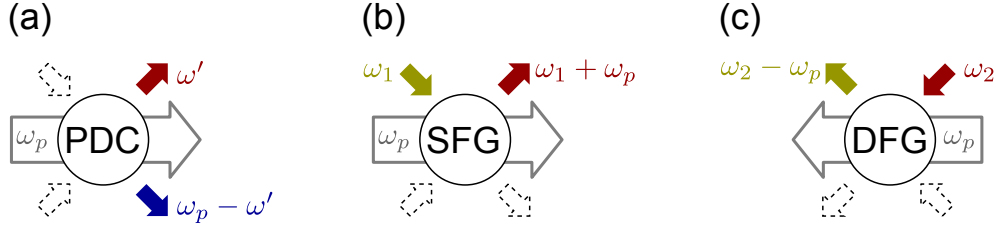}
	\caption{Schematics diagrams illustrating two different flavours of 
	second-order nonlinear processes. All processes are pumped by a classical
	undepleted field at a frequency $\omega_p$, the dashed arrows indicate
	vacuum modes. (a) In a PDC process two photons get created and the
	evolution operator for this process is a two-mode squeezing operator. 
	(b), (c) In a
	SFG or DFG process one incoming photon is annihilated and and outgoing
	photon at another frequency is created. Assuming that the frequencies
	involved in the two processes are equal, one can readily see that SFG
	and DFG are similar, yet reversed processes which is indicated in the
	schematics by the different directions the arrows point to. The 
	corresponding evolution operator for these processes is equivalent to
	a beamsplitter. For further details see text.}
	\label{figure_2}
\end{figure}
We identify $\theta$ with the beamsplitter angle which depends on the pump 
power and the strength of the nonlinear interaction. This will be discussed
later in great detail. We interpret this $\chitwo$-process as a beamsplitter
which diverts optical beams into different frequency output ports depending
on their initial frequency. Note that in
single-photon quantum optics this Hamiltonian describes SFG as well as DFG.
In classical nonlinear optics, however, DFG is understood as a stimulated process. The 
bright pump field has the highest frequency and the process is seeded with 
a weak input field which gets enhanced through continuous conversion of pump 
photons. In that case similar operator transformations as for PDC are valid
and the process could also be interpreted as seeded PDC.
In contrast, we assume a single (or few) photon input state, which has the 
highest frequency and the "seed" field is the bright field. By pinning the
pump field to a fixed value in our process model, we exclude that 
stimulation can occur and the process becomes formally equivalent to SFG.
Note that the usual no-pump-\emph{depletion} approximation 
($\partial E_\mathrm{p}/\partial z = 0$) needs to be interpreted as a 
no-pump-\emph{enhancement} approximation in this case. These findings are
schematically depicted in figure \ref{figure_2}.
We move on to the derivation of the interaction Hamiltonian
for quantum mechanical frequency conversion inside an optical waveguide.
%
%
%
\section{Quantitative derivation of the SFG and DFG interaction Hamiltonian}
\label{sec:hamiltonian}
\subsection{Spatial mode considerations in a monochromatic approach}
The interaction Hamiltonian of a frequency
conversion process can be expressed as
\begin{equation}
	\hint = -d_{\mathrm{eff}}\ezero\myint{^3r}E_\mathrm{p}(\vec{r}, t)
	\hat{E}^{(+)}_\mathrm{i}(\vec{r},t)
	\hat{E}_\mathrm{o}^{(-)}(\vec{r},t)+\hc,
	\label{eq:basic_hamiltonian_conv}
\end{equation}
where $d_{\mathrm{eff}}$ denotes the effective nonlinearity, $E_\mathrm{p}$ 
is the classical pump field and $\hat{E}^{(+)}_\mathrm{i}$ and 
$\hat{E}_\mathrm{o}^{(-)}$
denote the operator expressions for the
input signal and the converted 
output, respectively. As the interaction happens inside a nonlinear optical
waveguide, the propagation of the fields is restricted to one direction, 
which is given by the waveguide axis and which we define as
$z$-direction. The ultrafast pump field then reads
\begin{equation}
	E_\mathrm{p}(\vec{r}, t) = A_\mathrm{p}f_\mathrm{p}(x,y)\myint{
	\omega_\mathrm{p}}\alpha(\omega_\mathrm{p})\rme^{-\rmi
	\omega_\mathrm{p}t+\rmi\beta_\mathrm{p}z}.
	\label{eq:pump_field}
\end{equation}
Here, $\alpha(\omega_\mathrm{p})$ is the normalized spectral amplitude of the
pump. The function $f_\mathrm{p}(x,y)$ 
describes the transverse spatial distribution of the pump field with
$\myint{^2r}|f_\mathrm{p}(x,y)|^2=1$ and $\beta_\mathrm{p}$ is the 
propagation constant of the corresponding transverse mode. By requiring that
the area integration over the field intensity $I=\frac{1}{2}c\,n_\mathrm{p}
\ezero|E|^2$, where $n_\mathrm{p}$ denotes the refractive index at the pump
frequency, corresponds to a power, we find that the amplitude $A_\mathrm{p}$ is 
related to the average pump-pulse peak power $P_\mathrm{p}$ by
\begin{equation}
	A_\mathrm{p} = \lrsqrt{\frac{2 P_\mathrm{p}}{c\,\varepsilon_0 
	n_\mathrm{p}(\omega_\mathrm{p})|\myint{\omega_\mathrm{p}}\alpha(
	\omega_\mathrm{p})|^2}}.
\end{equation}
We implicitly make use of the slowly-varying envelope approximation
in this calculations, which is valid as we consider only pulses with 
$\Delta\omega\ll\omega_0$. This also means that we can neglect the frequency
dependence of the propagation constant $\beta_\mathrm{p}$ in  
\eref{eq:pump_field}.

To derive expressions for the quantized fields in a nonlinear
waveguide we start from the electric field operator for a propagating field
in a dielectric with finite cross-section area $\mathcal{A}$, given in 
\cite{Blow1990}. Note that we assume the wavevector components 
$k_x$ and $k_y$ of the  quantum field to have fixed, finite values.
\begin{equation}
	\fl\hat{E}^{(+)}(x,y,z,t) = \rmi\myint{\omega}\lrsqrt{\frac{\hbar\omega}
	{4\pi\ezero c\,n(\omega)\mathcal{A}}}\hat{a}(\omega)\rme^{-\rmi\omega t +
	\rmi(k_xx + k_yy + k_zz)}.
	\label{eq:loudon_efield}
\end{equation}
In a nonlinear waveguide with field propagation along $z$-direction, 
the solution of the Helmholtz equation yields a discrete spectrum of
valid propagation constants $\beta_{mn} = k_z^{(mn)}$ and a 
set of allowed, localized transverse modes $\{f_{mn}(x,y)\}$, determined 
by the boundary conditions of the guiding geometry. The indices $m$ and
$n$ denote the order of the transverse mode in $x$- and $y$-direction.
Each $\beta_{mn}$ corresponds to exactly one mode and,
in the case of a symmetric situation, the $\beta_{mn}$ for corresponding
modes (e.g. $\beta_{01}$ and $\beta_{10}$) become degenerate.
For better readability we only use one index $m$ to describe the modes.
The spatial localization of the fields implies continuous
spectra of the individual wavevector components $k_x^{(m)}$ and $k_y^{(m)}$,
given by $\tilde{f}_m(k_x,k_y) = \mathcal{FT}\left[f_m(x,y)\right]$\footnote{
We deploy the symmetric definition of the Fourier transform that is
$\tilde{f}(k)=\frac{1}{\sqrt{2\pi}}\myint{x}f(x)e^{-\rmi kx}$ and
accordingly $f(x)=\frac{1}{\sqrt{2\pi}}\myint{k}\tilde{f}(k)e^{\rmi kx}$.
}. 
The electric field inside a nonlinear waveguide is accordingly comprised of a 
superposition of quantum fields of the form given in \eref{eq:loudon_efield}, 
each corresponding to a single propagation direction.
Summing over the transverse waveguide modes and integrating over the
$k_x^{(m)}$ and $k_y^{(m)}$ we find:
\begin{eqnarray}
	\fl\hat{E}^{(+)}(x,y,z,t) =\rmi\sum_m\myint{\omega}\lrsqrt{\frac{
	\hbar\omega}{4\pi\ezero c\,n_m(\omega)}}\hat{a}_m(\omega)\rme^{-\rmi\omega
	t + \rmi\beta_mz}\times\nonumber\\
	\times\myint{k_x^{(m)}}dk_y^{(m)}\tilde{f}_m(k_x,k_y)
	\rme^{\rmi k_x^{(m)}x + \rmi k_y^{(m)}y}
\end{eqnarray}
This can -- due to the Fourier relationship between position and momentum --
be written as
\begin{equation}
	\fl\hat{E}^{(+)}(x,y,z,t) = \rmi\sum_m2\pi f_m(x,y)\myint{\omega}
	\lrsqrt{\frac{\hbar\omega}{4\pi\ezero c\,n_m(\omega)}}
	\hat{a}_m(\omega)\rme^{-\rmi\omega t + \rmi\beta_mz}.
\end{equation}
We would like to point out that we account for the cross-section area by the 
spatial distributions
$f_m(x,y)$ which are normalized such that $\int\rmd x\rmd y|f_m(x,y)|^2=1$ 
and which have units of inverse meters. Moreover we assume that within the 
frequency range of the considered fields the variation of the spatial 
properties is negligible, due to the narrowband approximation $\Delta\omega
\ll\omega_0$. We substitute the electric field operators into  
\eref{eq:basic_hamiltonian_conv} and rephrase the interaction Hamiltonian for 
single photon frequency conversion as
\begin{eqnarray}
	\fl\hat{H}_{int} = \frac{d_{\mathrm{eff}}\hbar\pi}
	{c}A_\mathrm{p}\sum_{l,m}\sqrt{\frac{\omega_\mathrm{i}
	\omega_\mathrm{o}}{n_{\mathrm{i},l}n_{\mathrm{o},m}}}\times\nonumber\\
	\times\int\rmd x\rmd y f_\mathrm{p}(x,y)f_{\mathrm{i},l}(x,y)
	f_{\mathrm{o},m}^*(x,y)\times\nonumber\\
	\times\myint{z}\rme^{\rmi(\beta_\mathrm{p}\pm\beta_{\mathrm{i},l}\mp
	\beta_{\mathrm{o},m})z}\times\nonumber\\
	\times\myint{\omega_\mathrm{p}}\rmd\omega_\mathrm{i}\rmd\omega_\mathrm{o}
	\alpha(\omega_\mathrm{p})\rme^{-\rmi(\omega_\mathrm{p}\pm\omega_{
	\mathrm{i}}\mp\omega_{\mathrm{o}})t}\aop_l(\omega_\mathrm{i})\cdag_m(
	\omega_\mathrm{o}) + \hc,
	\label{eq:hamiltonian}
\end{eqnarray}
where we discriminate between SFG and DFG. Here labels i and o denote input
and output fields, whereas the indices $l$ and $m$ describe the transverse
spatial modes of input and output field, respectively.


Now we move on to the calculation of the time evolution of the input quantum
state during the conversion process. Note that the interaction of the Hamilton
operator of equation \eref{eq:hamiltonian} is time-dependent and thus the
exact solution has to take into account time-ordering effects. Here
we present an approximate solution which neglects time-ordering effects, 
in order to emphasize the conceptual structure and to illustrate the main idea.
In \ref{sec:time-ordering} we validate this approach by comparing
the approximate solution with rigorous calculations we performed. 
We find that the shape of the mode functions does not change
significantly when taking into account time-ordering, but the maximum conversion
efficiency drops to 90\%. Still, these findings confirm that the
analytical solution leads to reasonable results and can safely be applied.
Hence we write the time evolution of the quantum state during the conversion
process
\begin{equation}
	\ket{\psi}_{out} = \hat{U}(t)\ket{\psi}_0 = 
	\exp\left(-\frac{\rmi}{\hbar}\myint{t}\hint(t)\right)\ket{\psi}_0.
\end{equation}
Thus, we need to perform the time integration of the interaction Hamiltonian
given in \eref{eq:hamiltonian}. This is a well-known procedure
discussed for PDC in great detail in \cite{URen2005}. We only present the
result here as the calculation, including the waveguide aspects, 
is straightforward.
\begin{eqnarray}
	\fl\myint{t}\hint(t) = \frac{2d_{\mathrm{eff}}\hbar\pi^2}{c}A_\mathrm{p}L
	\sum_{l,m}\sqrt{\frac{\omega_\mathrm{i}\omega_\mathrm{o}}{
	n_{\mathrm{i},l}n_{\mathrm{o},m}}}\frac{1}{\sqrt{A^{(eff)}_{l,m}}}\times
	\nonumber\\
	\times\myint{\omega_\mathrm{i}}\rmd\omega_\mathrm{o}\alpha(\omega_{
	\mathrm{io}})\phi_{l,m}(\omega_\mathrm{i},\omega_\mathrm{o})\aop_l(
	\omega_\mathrm{i})\cdag_m(\omega_\mathrm{o})+\hc
	\label{eq:output_state}
\end{eqnarray}
Here, $L$ is the length of the nonlinear waveguide. 
The function $\alpha(\omega_{\mathrm{io}})$ is the spectral pump distribution
defined as $\alpha(\omega_\mathrm{o}-\omega_\mathrm{i})$ for SFG and 
$\alpha(\omega_\mathrm{i}-\omega_\mathrm{o})$ for DFG, respectively, whereas
the function $\phi_{l,m}(\omega_\mathrm{i},\omega_\mathrm{o})$ 
characterizes the phasematching and is given by
\begin{equation}
	\phi_{l,m}(\omega_\mathrm{i},\omega_\mathrm{o}) = \mathrm{sinc}\left(\
	\frac{\Delta\beta_{l,m}L}{2}\right)\approx\exp\left[-0.193\cdot
	\left(\frac{\Delta\beta_{l,m}L}{2}\right)^2\right].
\end{equation}
The expression $\Delta\beta_{l,m}$ describes the phase-mismatch of the propagation 
constants
and evaluates to $\Delta\beta_{l,m}=\beta_\mathrm{p}+\beta_{\mathrm{i},l}-
\beta_{\mathrm{o},m}-\frac{2\pi}{\Lambda}$ for SFG and $\Delta\beta_{l,m}=
\beta_\mathrm{p}-\beta_{\mathrm{i},l}+\beta_{\mathrm{o},m}-\frac{2\pi}{
\Lambda}$ for DFG, respectively. Finally, $\Lambda$ is an optional poling period for
quasi-phasematching inside the waveguide. Following the usual conventions we
define an \emph{effective interaction area} $A^{(eff)}_{l,m}$:
\begin{equation}
	\frac{1}{A^{(eff)}_{l,m}} := \left[\myint{x\rmd y}f_\mathrm{p}(x,y)
	f_{\mathrm{i},l}(x,y)f_{\mathrm{o},m}^*(x,y)\right]^2.
\end{equation}
Note that this should not be mistaken as a geometric area defined for 
instance by the waveguide cross section. Instead it describes the overlap of 
the transverse spatial modes of the three interacting fields inside the 
nonlinear waveguide. This result also implies that simply using a smaller 
waveguide -- while not changing the modal overlap characteristics -- will not alter
$A^{(eff)}$ and will therefore not have any impact on the efficiencies of the
processes. The product of pump distribution and phasematching function is
conveniently defined as joint spectral distribution function
\begin{equation}
	G_{l,m}(\omega_\mathrm{i},\omega_\mathrm{o}) = \frac{1}{N_{l,m}}
	\alpha(\omega_{\mathrm{io}})\phi_{l,m}(\omega_\mathrm{i},
	\omega_\mathrm{o}),
\end{equation}
which describes the mapping between input and output frequencies for a 
specific pair of spatial modes $l,m$. The normalization factor $N_{l,m}$ reads 
$\left(\myint{\omega_\mathrm{i}}\rmd\omega_\mathrm{o}|\alpha(_{\mathrm{io}})
\phi_{l,m}(\omega_\mathrm{i},\omega_\mathrm{o})|^2\right)^{1/2}$. 
\subsection{Broadband pulse mode picture}
The description derived so far has been in terms of
monochromatic creation and annihilation operators. However, since we 
concentrate on $\chitwo$-interactions between ultrafast pulses,
a much more natural approach is to consider broadband pulse modes. 
A suitable pulse-mode basis is found by applying a Schmidt decomposition
to the joint spectral distribution function:
\begin{equation}
	G_{l,m}(\omega_\mathrm{i},\omega_\mathrm{o}) = \sum_{j}\kappa^{(l,m)}_j
	\varphi^{(l,m)}_j(\omega_\mathrm{i})\psi^{(l,m)}_j(\omega_\mathrm{o}).
	\label{eq:schmidt_decomposition}
\end{equation}
Equation \eref{eq:schmidt_decomposition} yields two correlated sets of orthonormal
broadband pulse-mode functions $\{\varphi^{(l,m)}(\omega_\mathrm{i})\}$ and
$\{\psi^{(l,m)}(\omega_\mathrm{o})\}$. The diagonal values $\kappa_j^{(l,m)}$ 
are the real and positive Schmidt coefficients and satisfy $\sum_j(
\kappa_j^{(l,m)})^2=1$. It is well known for PDC that the basis sets of 
the Schmidt decomposition, and thus the modal structure of the photons, are 
uniquely defined \cite{Martinelli2003}. The same argument can also be applied
here in the context of SFG and DFG. As for PDC we define
broadband creation and annihilation operators:
\begin{eqnarray}
	\hat{A}_{j,l,m} = \myint{\omega_\mathrm{i}}\varphi^{(l,m)}_j(
	\omega_\mathrm{i})\aop_l(\omega_\mathrm{i}),\\
	\hat{C}_{j,l,m} = \myint{\omega_\mathrm{o}}\psi^{(l,m)}_j(
	\omega_\mathrm{o})\cop_m(\omega_\mathrm{o}).
\end{eqnarray}
Substituting those, we rewrite the expression for the time-integrated 
interaction Hamiltonian from \eref{eq:output_state}, and
arrive at the broadband pulse-mode picture
\begin{eqnarray}
	\fl\myint{t}\hint(t) = \frac{2d_{\mathrm{eff}}\hbar\pi^2}{c}A_\mathrm{p}L
	\sum_{l,m}\sqrt{\frac{\omega_\mathrm{i}\omega_\mathrm{o}}{n_{
	\mathrm{i},l}n_{\mathrm{o},m}}}\frac{N_{l,m}}{\sqrt{A^{(eff)}_{l,m}}}
	\times\nonumber\\
	\times\sum_j\left(\kappa^{(l,m)}_j\hat{A}_{j,l,m}\hat{C}^{\dagger}_{
	j,l,m}+\hc\right)=\nonumber\\
	= \hbar\sum_{l,m}\sum_j\theta_{j,l,m}\left(\hat{A}_{j,l,m}
	\hat{C}^{\dagger}_{j,l,m} + \hat{A}^{\dagger}_{j,l,m}\hat{C}_{j,l,m}
	\right).
	\label{eq:output_state_2}
\end{eqnarray}
By introducing the effective coupling constant $\theta_{j,l,m}$ in 
\eref{eq:output_state_2}, we reveal the simple beamsplitter structure of the
Hamiltonian \cite{Raymer2010}, as already announced in 
\eref{eq:intro_beamsplitter}. In contrast to a conventional 
beamsplitter however, this Hamiltonian does not couple two $k$-modes 
(or beam paths), but rather two broadband pulse modes 
$\hat{A}_{j,l,m}$ and $\hat{C}_{j,l,m}$ at different frequencies! 
This is a unique feature of ultrafast frequency conversion 
processes and makes them ideal candidates for the implementation of the
QPG and QPS.

%% file: part_3_arxiv_2.tex
\section{Pushing towards applications}
\label{sec:application}
\subsection{General non-engineered SFG and DFG}
\begin{figure}
	\centering
	\includegraphics[width=\linewidth]{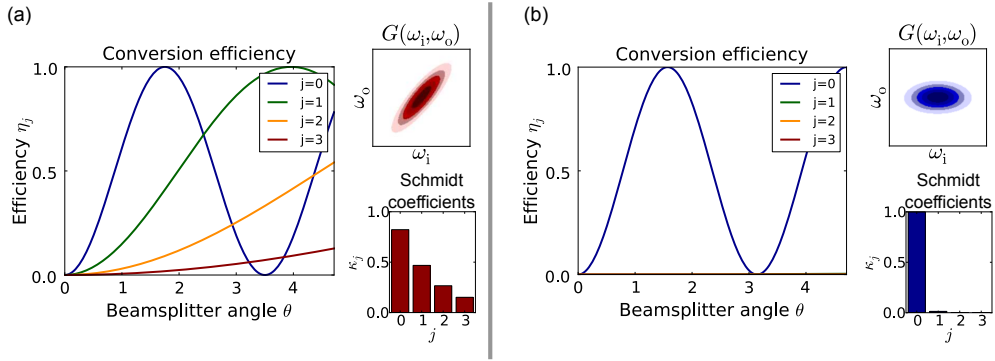}
	\caption{Conversion efficiencies $\eta_j$, joint spectral distribution
	functions $G(\omega_\mathrm{i}, \omega_\mathrm{o})$ and 
	Schmidt coefficients $\kappa_j$ for the first four pairs of 
	pulse modes ($j=0\dots3$). (a)
	Non-engineered process with several $\kappa_j\neq0$. For a given overall
	beamsplitter angle $\theta$, all modes are converted to a certain extent
	given by $\eta_j = \sin(\kappa_j\cdot\theta)^2$. However,
	an overall unit efficiency can generally not be accomplished. (b) 
	Source-engineered process with one predominant $\kappa_j\approx1$. 
	By choosing an appropriate $\theta$, pulse mode $\varphi(\omega_\mathrm{
	i})$ can be converted into pulse mode $\psi(\omega_\mathrm{o})$ with unit
	efficiency, allowing for QPG operation.}
	\label{figure_3}
\end{figure}

QPG and QPS are unique in their single-mode operation on broadband
pulse modes. In this section we discuss the implementation of 
genuine QPG or QPS in a feasible experimental setup. We restrict
the analysis to only one pair of transverse spatial modes $(l,m)$, which
simplifies the notation but does not change the underlying physics. In the
experimental setting the selection of one spatial mode can be accomplished 
by broadband spectral filtering \cite{Mosley2009}. In this case the
time-integrated, effective SFG- and DFG-Hamiltonian from \eref{eq:output_state_2}
reads
\begin{equation}
	\myint{t}\hint(t) = \hbar\sum_j\theta_j\left(
	\hat{A}_j\hat{C}^{\dagger}_j + \hat{A}^{\dagger}_j\hat{C}_j\right),
	\label{eq:hamiltonian_applications}
\end{equation}
with the broadband operators defined as
\begin{eqnarray}
	\hat{A}_j = \myint{\omega_\mathrm{i}}\varphi_j(\omega_\mathrm{i})\hat{a}
	(\omega_\mathrm{i}),\\
	\hat{C}_j = \myint{\omega_\mathrm{o}}\psi_j(\omega_\mathrm{o})\hat{c}
	(\omega_\mathrm{o}).
\end{eqnarray}
The complete, orthonormal function sets $\{\varphi_j(\omega_\mathrm{i})\}$ and $\{\psi_j(\omega_\mathrm{o})\}$
represent the intrinsic pulse-mode structure of the SFG- or DFG-process, 
obtained from the Schmidt-decomposition of the joint spectral distribution 
function $G(\omega_\mathrm{i},\omega_\mathrm{o})=\alpha(\omega_{\mathrm{io}})
\phi(\omega_\mathrm{i},\omega_\mathrm{o})$. On the one hand these are determined
by the pump pulse characteristics, but on the other hand they also critically depend
on the nonlinear waveguide's material and modal dispersion properties. We have already
stressed the formal equivalence between the expression from 
\eref{eq:hamiltonian_applications} and a sum of optical beamsplitter 
Hamiltonians. Hence the linear transformation for the broadband operators can
be readily written as
\begin{equation}
	\hat{A}_j\rightarrow\cos(\theta_j)\hat{A}_j-\rmi\sin(\theta_j)\hat{C}_j,
\end{equation}
corresponding to a pulse mode conversion between $\varphi_j(\omega_\mathrm{i})$
and $\psi_j(\omega_\mathrm{o})$ with efficiency $\eta_j = \sin^2(
\theta_j)$. According to \eref{eq:output_state_2} the coupling constant 
$\theta_j$ is given by
\begin{equation}
	\theta_j = \kappa_j\cdot\frac{2d_{\mathrm{eff}}\pi^2LN}{c}\sqrt{\frac{
	2\omega_\mathrm{i}\omega_\mathrm{o}}{c\,\ezero n_\mathrm{p}n_\mathrm{i}
	n_\mathrm{o}|\myint{\omega_\mathrm{p}}\alpha(\omega_\mathrm{p})|^2}}
	\sqrt{\frac{P_\mathrm{p}}{A^{(eff)}}}=\kappa_j\cdot\theta.
\end{equation}
Here, $\theta$ is an overall beamsplitter angle defined by the process parameters.
Its impact on the different modes $j$ is given by $\theta_j$, where, for
each mode, the overall beamsplitter angle is weighted with the corresponding
Schmidt coefficient $\kappa_j$.
In figure \ref{figure_3} (a) we illustrate a general, non-engineered SFG.
We show the joint spectral distribution function
$G(\omega_\mathrm{i},\omega_\mathrm{o})$ as well as the Schmidt coefficients
$\kappa_j$ for the first four pairs of pulse modes and 
plot the conversion efficiencies $\eta_j$ versus the beamsplitter angle
$\theta$. It is obvious that, for any given value of $\theta$, all pulse modes
with $\kappa_j\neq0$ are converted to a certain extent. Yet, in general
neither single-mode operation is achievable, nor can conversion with 
$\eta_j=1$ for different pulse modes simultaneously be accomplished. We
note an exception to this rule: Under certain conditions (e.g. a 
cw pump) input and output modes are perfectly correlated. Then all 
$\kappa_j$ share the same value and all modes are converted with the same
efficiency. The process is then highly multimode but the overall efficiency
can reach unity for high pump powers. 

\subsection{Source-engineered SFG and DFG -- towards genuine QPG and QPS}
We have shown that SFG and DFG in general are multimode processes. But
for QPG and QPS we require single-mode operation in order to avoid signal degradation
introduced by vacuum contributions and to achieve unit efficiency. Reducing the intrinsic
pulse-mode structure of a $\chitwo$-nonlinear process to only one pair of 
modes has been extensively studied in PDC, where source-engineering led to the 
desired results \cite{URen2005,Mosley2008,Wong2008}. Experimentally this
is accomplished by group-velocity matching inside the nonlinear medium. If
pump and either signal or idler share the same group velocity, the 
phasematching function becomes parallel to one of the axes when plotted in
an $(\omega_\mathrm{s},\omega_\mathrm{i})$-diagram. Then, the Schmidt 
decomposition yields -- given that the process is pumped by an ultrafast 
pump -- only one pair of pulse modes which is excited with unit efficiency.

We transfer this insight to our analysis of SFG and DFG and employ it for
spectral engineering of the conversion. The time-integrated Hamiltonian
from \eref{eq:hamiltonian_applications} for our special case reduces to
\begin{equation}
	\myint{t}\hint=\hbar\theta(\hat{A}\hat{C}^{\dagger}+\hat{A}^{\dagger}
	\hat{C}),	
\end{equation}
and can be interpreted as a beamsplitter operating on only one pair of 
pulse modes $\varphi(\omega_\mathrm{i})$ and $\psi(\omega_\mathrm{o})$.
As an example, we show an engineered
case in figure \ref{figure_3} (b) where pump and input signal are 
group-velocity matched. Note that, in contrast to the previous
non-engineered case, the joint spectral distribution function now shows 
no spectral correlations between input and output frequencies that is it 
is oriented
along the axes of the diagram. As in PDC this is a direct consequence of
the horizontally oriented phasematching function and thus of the 
group-velocity matching. The distribution of the $\kappa_j$ reveals that only
one coefficient $\kappa_0$ differs significantly from zero. This is also
reflected in the plot of the conversion efficiencies. Only one single pulse 
mode is addressed and, by choosing $\theta=\frac{\pi}{2}$, 
converted with unit efficiency and no vacuum is coupled into the signal beam. Hence we find that, by group-velocity
matching pump and either input or output, we can achieve genuine single-mode
operation and therefore implement QPG and QPS. Note that the data in figure
\ref{figure_3} have been calculated using our modeling and realistic
experimental parameters, which are specified in section 
\ref{sec:performance}. 

\begin{figure}
	\includegraphics[width=\linewidth]{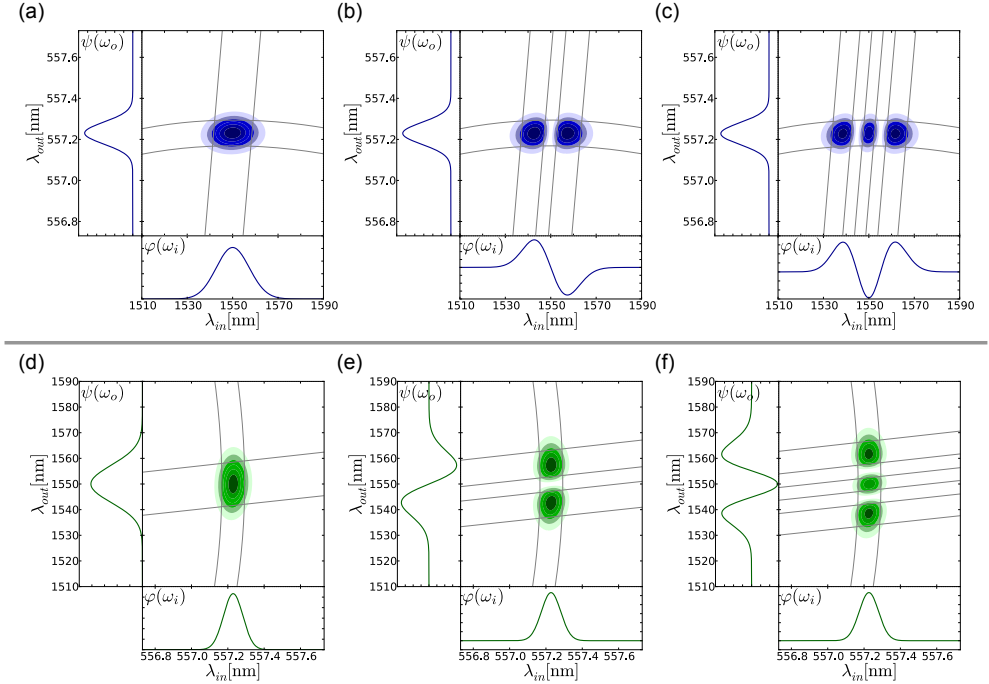}
	\caption{Joint spectral distribution, pump and phasematching function
	as well as pulse modes $\varphi(\omega_\mathrm{i})$ and $\psi(
	\omega_\mathrm{o})$ for QPG (a)-(c) and QPS (d)-(f). Note that the 
	functions are plotted against wavelengths for reasons of convenience.
	In a QPG, the intrinsic pulse mode $\varphi(\omega_\mathrm{i})$ can be
	manipulated	by shaping the bright gating pulse, whereas pulse mode $\psi(
	\omega_\mathrm{o})$	is fixed by the phasematching function. Therefore,
	arbitrary input	modes are mapped to the same output mode, allowing for
	interference of formerly orthogonal modes. Contrary to that, in a QPS,
	shaping the bright pulse defines the output pulse mode $\psi(
	\omega_\mathrm{o})$. The input pulse mode is now defined by the 
	phasematching function. Thus, an arbitrary mode can be generated from 
	an input which is matched to $\varphi(\omega_\mathrm{i})$. 
	The data presented here is calculated using realistic experimental 
	parameters, specified in section \ref{sec:performance}.}
	\label{figure_4}
\end{figure}

Knowing how to achieve single-mode operation of SFG and DFG, the next
step is to investigate how we can exact complete control over the pulse modes
$\varphi(\omega_\mathrm{i})$ and $\psi(\omega_\mathrm{o})$. A QPG selects a
specific pulse mode from an input state and a QPS generates an arbitrary 
pulse mode from a Gaussian input mode. Hence, for QPG we require control
over $\varphi(\omega_\mathrm{i})$, whereas for QPS we require shaping of 
$\psi(\omega_\mathrm{o})$, respectively. In figure \ref{figure_4} (a)-(c) we
consider QPG. Shown are the phasematching and pump functions as well as the
resulting joint spectral distribution function. Note that the axes are
given wavelength units rather than frequency for convenience.
We find that the output mode $\psi(
\omega_\mathrm{o})$ is defined solely by the phasematching. We
performed calculations for three different spectral shapes of the pump and it is
obvious that the input mode $\varphi(\omega_\mathrm{i})$ has the form of the
respective pump mode. Thus, in a QPG, spectrally shaping the bright gating pulse leads
to the selection of an arbitrary pulse mode. In contrast, figure 
\ref{figure_4} (d)-(f) illustrate the situation for QPS. Now, pump and output
are group-velocity matched, causing a vertical phasematching function.
Again we calculated three different spectral pump distributions. We find
that the spectral shape of the pump now defines the output mode $\psi(
\omega_\mathrm{o})$ of the QPS and the input mode $\varphi(
\omega_\mathrm{i})$ is fixed by the phasematching. Summing up these findings,
we end up with the following correspondences:
\begin{eqnarray}
	\alpha(\omega_\mathrm{p})&\rightarrow \varphi(\omega_\mathrm{i}),
	\quad\phi(\omega_\mathrm{i},\omega_\mathrm{o})&\rightarrow \psi(
	\omega_\mathrm{o})\quad\mathrm{for}\:\mathrm{QPG},\\
	\phi(\omega_\mathrm{i},\omega_\mathrm{o})&\rightarrow \varphi(
	\omega_\mathrm{i}),\quad\alpha(\omega_\mathrm{p})&\rightarrow \psi(
	\omega_\mathrm{o})\quad\mathrm{for}\:\mathrm{QPS}.
\end{eqnarray}
We demonstrated that we can achieve complete control over the required pulse mode of the
QPG or QPS by shaping the bright gating pulse or shaping pulse, respectively.
Note however that we considered only the intrinsic modes of the QPG and QPS,
which do not necessarily have to coincide with the pulse-mode structure of
an input signal.

\subsection{Mode matching a QPG or a QPS}
Given a specific QPG or QPS, the input signal's pulse-mode structure must
coincide with the pulse-modes $\{\varphi(\omega_\mathrm{i})\}$ accepted by
the device, in order to guarantee mode selectivity and high conversion
efficiency. We first discuss
this for the QPG. We have shown that the QPG can be easily adapted to
a wide range of input signals by spectrally shaping the coherent gating pulse. 
The output mode $\psi(\omega_\mathrm{o})$ is solely defined by the 
phasematching function and is independent from
the pump pulse shape. It can typically be
approximated by a Gaussian spectrum\cite{URen2005}. Hence any selected
mode from an input state is mapped to the same output mode.

\begin{figure}
	\centering
	\includegraphics[width=\linewidth]{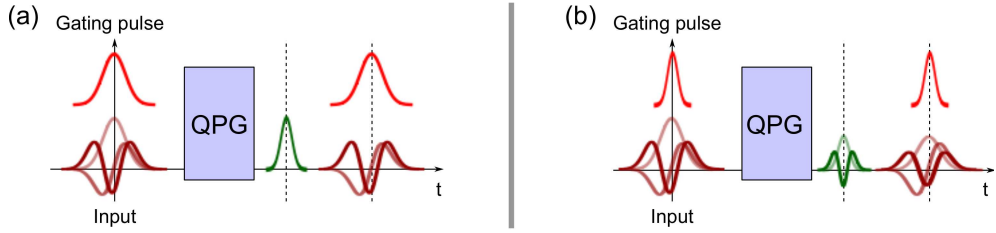}
	\caption{(a) Mode-matched QPG. The bright gating pulse has the
	same duration as the input signal, leading to the matching of the
	input pulse-mode structure and $\varphi(\omega_\mathrm{i})$ of the QPG.
	Only a single pulse mode of the input -- the one which overlaps with 
	$\varphi(\omega_\mathrm{i})$ -- is selected and converted with unit 
	efficiency. (b) Mode-mismatch in a QPG. The gating pulse and input signal 
	have different durations, leading to an overlap of $\varphi(
	\omega_\mathrm{i})$ with all input signal modes with the same parity. 
	Hence, all of those modes are selected and converted to a certain degree.
	QPG-operation is then not possible.}
	\label{figure_5}
\end{figure}

In figure \ref{figure_5} (a) we illustrate this situation. Input
signal and gating pulse share the same duration and the QPG is mode-matched
to the input. Only the desired mode from the input signal is selected
and converted with unit efficiency. In contrast, figure \ref{figure_5} (b)
demonstrates the impact of a mode mismatch on the QPG operation. The gating
pulse duration significantly differs from the input pulse duration
and the intrinsic QPG pulse mode
$\varphi(\omega_\mathrm{i})$ overlaps with all signal modes of the same 
parity. We end up with a case similar to multimode SFG, with the only 
difference that the diverse conversion efficiencies for the modes are due
to the different overlaps between $\varphi(\omega_\mathrm{i})$ and the
corresponding signal-state modes. Because an overall efficiency of unity 
can not be achieved here and the process is not 
mode-selective anymore, it becomes clear
that careful mode-machting is vital for a successful QPG implementation.

The situation is different for QPS: The
phasematching function is vertical in the 
$(\omega_\mathrm{i},\omega_\mathrm{o})$-plane (compare
figure \ref{figure_4} (c)-(f)) and the input mode $\varphi(
\omega_\mathrm{i})$ is now defined by the phasematching function alone.
The QPS accepts only Gaussian input modes which are matched to $\varphi(
\omega_\mathrm{i})$. However, shaping the bright pulse allows for defining
the output mode $\psi(\omega_\mathrm{o})$, rendering it possible to generate
any pulse mode from an input pulse with a Gaussian spectrum. 
If the input state is not matched
to $\varphi(\omega_\mathrm{i})$, this does not change the spectral form of
the output pulse. The drawback is that not the complete
input gets converted and vacuum contributions are introduced.

We note that the same physical nonlinear waveguide device could be used
as QPG or QPS, depending on whether SFG or DFG is implemented:
this shows that QPG and QPG can be seen as reverse operations of each other.
The results illustrate that QPG and
QPS are versatile tools which can be easily adapted to a large range of
input and output states, making them highly flexible and appealing for
many applications.

\section{Performance of QPG and QPS considering realistic 
experimental parameters}
\label{sec:performance}
\begin{figure}
	\centering
	\includegraphics[width=.5\linewidth]{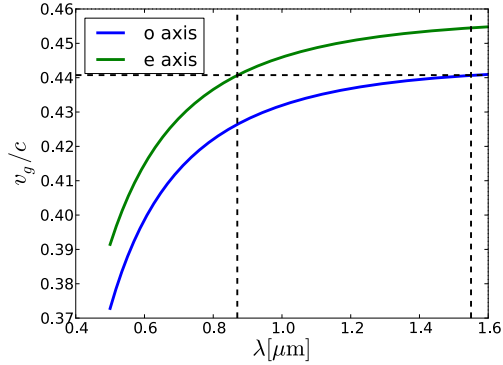}
	\caption{Group velocities of the fundamental waveguide modes in
	a Ti-indiffused PPLN waveguide. For a 1550nm input oriented along
	the ordinary axis, an extraordinarily polarized pump pulse 
	centered around 870nm is group-velocity matched. 
	For a QPG, this consequently leads to an output
	at 557nm which has to be oriented along the ordinary axis. It becomes 
	obvious that for a wide range of input signal wavelengths,
	group-velocity matched gating pulses can be found which still 
	satisfy feasible experimental parameters.}
	\label{figure_6}
\end{figure}

\begin{figure}
	\includegraphics[width=\linewidth]{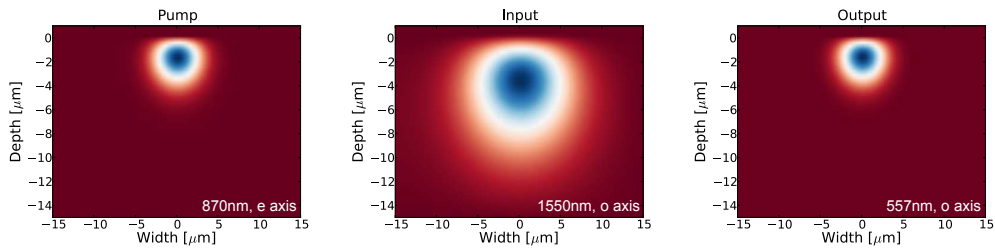}
	\caption{Transverse spatial mode profiles of the fundamental waveguide
	modes in a Ti-indiffused PPLN waveguide. The mode profiles were 
	calculated using a finite elements method. The overlap between the pump and 
	the output 
	exceeds 99\%, because the guiding for ordinarily polarized fields is not 
	as strongly pronounced as for extraordinarily polarized fields.}
	\label{figure_7}
\end{figure}

We conclude our analysis demonstrating the experimental feasibility of a 
QPG and 
derive, with the help of the theoretical model outlined in 
sections \ref{sec:transformations} to \ref{sec:application}, an expression for
the pump power for maximally efficient operation. The results of the
calculation apply to QPS as well, since both devices can be implemented in the
same nonlinear waveguide. SFG phasematching implies an existing DFG phasematching,
only the roles of input and output field are interchanged. The coupling constant
$\theta$ is the same for both processes. The bright pulse, used
as gating or shaping pulse depending on the application, will be called 
pump in this paragraph for 
ease of reading. The input field is at 1550nm and the third field is
referred to as output. 

As a key point for the experimental setup we require that it 
can be operated at 1550nm. The constraint of group-velocity matching 
determines the pump wavelength as soon as the input wavelength gets
fixed, which in turn then also defines the output wavelengths due to
energy conservation.
We assume that the conversion takes place in a 
Ti-indiffused PPLN waveguide with a length of $L=10mm$ and at a 
temperature of 
$T=190^{\circ}$C to prevent the impact of photorefraction. Effective
Sellmeier equations for the three participating fields were obtained
by calculating the effective refractive indices of ordinary and  
extraordinary polarized fields with a finite-element method 
described in \cite{Strake1988}. The effective equations were then 
fitted against the calculated values. Note that the following 
calculations are based on these effective Sellmeier equations.

In figure 
\ref{figure_6} we plot the group velocities for the ordinary and extraordinary
crystal axes and assume for our modeling that all fields propagate in the fundamental
transverse waveguide mode. If the input light is ordinarily (TE-) polarized with 
central wavelength of 1550nm, we find that the group-velocity matched pump
has to be extraordinarily (TM-) polarized and centered 
around 870nm. The ordinarily-polarized output is then at 557nm. 
From figure \ref{figure_6} we can clearly recognize that a group-velocity matched
pump can be found for any input, as long as the input is o-polarized. 
The effective refractive indices of the participating fields 
calculate to $n_\mathrm{p} = 2.18$, $n_\mathrm{i} = 2.21$ and $
n_\mathrm{o} = 2.32$ and we derive 
a periodic poling period of $\Lambda\approx4.28\mu$m required for 
quasi-phasematching inside the waveguide.

In figure \ref{figure_7} we plot the 
transverse spatial distributions of the input, pump and output
modes, also obtained with the finite-element method from \cite{Strake1988}.
From these we calculate the effective interaction area $A^{(eff)}
\approx64\mu\mathrm{m}^2$.
The conversion efficiency for a single-mode operation is $\eta = 
\sin^2(\theta)$ and the condition for unit efficiency can be specified by
\begin{eqnarray}
	\theta \stackrel{!}{=} \frac{\pi}{2},\\
	P_\mathrm{p} \stackrel{!}{=} \left(\frac{c}{4\pi d_{\mathrm{eff}}LN}
	\right)^2\frac{c\,\varepsilon_0n_\mathrm{p}n_\mathrm{i}n_\mathrm{o}
	|\myint{\omega_\mathrm{p}}\alpha(\omega_\mathrm{p})|^2
	A^{(eff)}}{2\omega_\mathrm{i}\omega_\mathrm{o}}.
\end{eqnarray}
Assuming an input pulse duration of roughly 300fs, we calculate a 
required pump peak 
power of $P_\mathrm{p}\approx22\mathrm{W}$ for optimal conversion efficiency. 
If a pump laser system with 
a repetition frequency of 76MHz is used, we obtain an average pump power
of $P_{\mathrm{av}}\approx0.5\mathrm{mW}$ inside the waveguide. 
This leads to required average pump powers of a few mW in front of 
the QPG or QPS, taking into account 
realistic waveguide coupling losses. These values are lower than
formerly reported pump powers for similar experiments 
\cite{Roussev2004, VanDevender2004, 
Langrock2005, Tanzilli2005, Rakher2010}, owing to the careful 
source-engineering we applied to our process. 
This grants a significant 
advantage over experiments without spectrally engineered SFG, even though 
we employ a cross-polarized process with an 
effective nonlinearity that is lower by an order of magnitude compared to 
a process
where all three fields are oriented along the extraordinary crystal axis.

%% file: part_4_arxiv_2.tex
\section{Conclusion and Outlook}
\label{sec:conclusion}
In conclusion, we presented a feasible way to achieve complete control over
the pulse-mode structure of ultrafast pulsed quantum states of light. We combined
findings from quantum state generation and techniques from state manipulation,
by applying spectral-source engineering and integrated optics to frequency
conversion of ultrafast single photons. We showed that single-mode ultrafast
sum- and difference-frequency generation in $\chitwo$-nonlinear materials
are possible and analyzed two
highly flexible and versatile devices, namely quantum pulse gate and quantum
pulse shaper. The QPG is based on ultrafast SFG and offers the possibility
to select arbitrary pulse modes from an ultrafast multimode input state. 
The selected mode gets converted with unit efficiency and is mapped onto
a Gaussian output mode. The residual mode structure of the input is left 
intact, allowing for cascaded operation to convert multiple modes. As 
all input modes are mapped onto the same output mode, interference of formerly
orthogonal states becomes possible. In contrast, the newly-introduced 
QPS is based on DFG and
implements the reverse operation of a QPG. 
It enables us to generate an arbitrary pulse form from a Gaussian input mode.
The output mode is defined by a bright shaping pulse, thus highly flexible
state preparation can be achieved. We have presented a quantitative analysis
of QPG and QPS and derived feasible experimental parameters with which the
proposed devices can be implemented, rendering them practical instead of
merely conceptual.

As a final remark we would like to point out that our analysis is in no way
constrained to single-photon states. Although we consider single-photon input
states, the introduced concepts can be generalized
to classical and non-classical multi-photon states. In this framework the 
use of QPG and
QPS provides the attractive opportunity to successively select and spatially
separate arbitrary pulse modes from a multimode input state while leaving
the residual beam intact. Employing a series of QPGs operating on the same
pulse mode in each arm of a multimode twin-beam squeezer source allows for 
a feasible implementation of non-Gaussian operations and thus constitutes
an important step towards the
realization of multimode continuous variable entanglement distillation. QPS on
the other hand can be used to synthesize multimode continuous variable Gaussian
states in a mode-by-mode fashion. The prepared states can then subsequently 
be transmitted as a bundle, since they do not interact with each other and
they all experience the same dispersion during transmission and therefore
stay orthogonal. This paves the way towards dense channel-multiplexing
in continuous variable quantum communication.

\section*{Acknowledgments}
\addcontentsline{toc}{section}{Acknowledgments}
We would like to thank Michael Raymer for invaluable discussions of
this work.

The research leading to these results has received funding from the 
European Community's Seventh Framework Programme FP7/2001-2013 under
grant agreement no. 248095 through the Integrated Project Q-ESSENCE.

	\section*{Appendix. Impact of time-ordering}
	\addcontentsline{toc}{section}{Appendix. Impact of time-ordering}
	\label{sec:time-ordering}
	Since the interaction of the Hamilton operator of equation 
	\eref{eq:hamiltonian} is time-dependent, it might be assumed that 
	time-ordering effects have a major impact on the intrinsic mode structure 
	of the process, in particular if a perturbative solution is not sufficient.
	This case is associated with unit conversion efficiency, needed for 
	perfect QPG and QPS operation. In our analysis we have nevertheless 
	deployed the approximate solution which neglects time-ordering.
	The impact of time-ordering on the process of ultrafast
	PDC has been thoroughly investigated in \cite{Wasilewski2006}, with the
	result that time-ordering mostly affects the amplitudes but not the 
	shapes of the intrinsic pulse modes. In \cite{Kielpinski2010}, the authors 
	actually study a three-wave mixing process and find in their numerical 
	simulation no major discrepancy with their analytical solution.
	This already indicates that, at least for low conversion efficiencies 
	where a perturbative solution is sufficient, time-ordering can be neglected.
	However, since we aim for conversion efficiencies of unity and cannot 
	conclude for sure that the above results remain valid in our case, we  
	performed rigorous numerical simulations which take into account all 
	time-ordering effects. Note that the results presented are all obtained for 
	the case of maximum conversion efficiency. Additionally, the simulated
	processes are the ones discussed in this work. That is, pump pulse
	and input pulse are group-velocity matched and the processes have
	decorrelated joint spectral distribution functions as is the case
	in figure \ref{figure_3} (b).
	
	\begin{figure}
		\centering
		\includegraphics[width=0.85\linewidth]{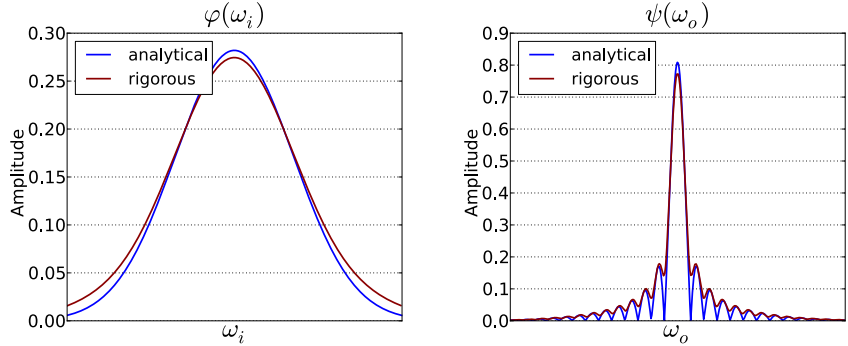}
		\caption{Input modes $\varphi(\omega_i)$ as well as output modes
		$\psi(\omega_o)$ of the considered processes, obtained with the
		analytical solution (blue) and the rigorous calculation (red),
		respectively. Obviously, time-ordering has only a small impact
		on the actual mode shape. This can simply be corrected for in our
		schemes by adjusting the spectrum of the bright pump pulse.
		Note that the oscillations in the output modes
		originate in the sinc function which describes the phasematching.}
		\label{figure_8}
	\end{figure}
	
\begin{figure}
		\centering
		\includegraphics[width=0.45\linewidth]{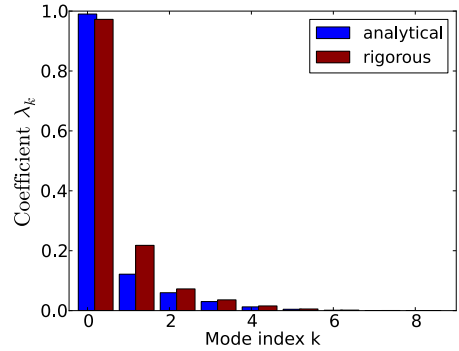}
		\caption{Schmidt coefficients $\lambda_k$ obtained from the analytical
		as well as the rigorous solution. Time-ordering causes a slight shift
		in the excitation of the different modes. In addition it leads to a
		drop in the maximum conversion efficiency to 90\%, as compared to 
		a unit conversion efficiency reached in the analytical solution.}
		\label{figure_9}
	\end{figure}

	Figure \ref{figure_8} depicts the analytical as well as the rigorous input
	modes $\varphi_\mathrm{ana}(\omega_i)$ and $\varphi_\mathrm{rig}(\omega_i)$
	and the corresponding output modes $\psi_\mathrm{ana}(\omega_o)$ and
	$\psi_\mathrm{rig}(\omega_o)$, respectively.
	It nicely illustrates that time-ordering has only a slight
	impact on the shape of the modes, as expected from \cite{Wasilewski2006}. The
	change in the modes can easily be compensated for in our proposed scheme, 
	by adjusting the spectrum of the bright gating pulse. Note that the
	oscillations in the output modes originate from the sinc function which describes
	the phasematching. These also cause the slight multi-modeness which can be
	seen in the Schmidt coefficients in figure \ref{figure_9}, where the first
	higher order mode is also excited with a certain probability. Comparing
	again analytical and rigorous solutions, we find that time-ordering slightly
	shifts the weights between the different modes. However, we want to note that
	no new modes occur in the process due to time-ordering. The main difference
	is found when considering the maximum conversion efficiency. It turns out
	that this value drops in the rigorous solution to 90\% instead of the
	unit efficiency obtained with the analytical approach.
	
	This behavior will be thoroughly 
	analyzed in \cite{Christ2011}. We want to stress here that all characteristics
	introduced by the side lobes of the sinc function can be washed out through careful
	design of the nonlinearity inside the waveguide as shown in \cite{Branczyk2011}.
	Therefore the device performance calculated here only represents a lower bound
	and might be increased in the future, for instance by implementing a Gaussian
	shaped phasematching function.

	In conclusion we find by comparing analytical and numerical solutions that
	the assumption that time-ordering can be neglected is, in fact, a rather
	good approximation, even for the cases of high conversion efficiencies 
	analyzed in this work.